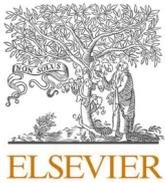
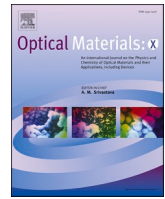

Invited Article

# (INVITED) Ultraviolet cross-luminescence in ternary chlorides of alkali and alkaline-earth metals

V. Vaněček [a,b], J. Páterek [a,b], R. Král [a], R. Kučerková [a], V. Babin [a], J. Rohlíček [a], R. Cala' [c,d], N. Kratochwil [c,e], E. Auffray [c], M. Nikl [a,*]

[a] *FZU, Institute of Physics of the Czech Academy of Sciences, Cukrovarnicka 10, Prague, Czech Republic*
[b] *FNSPE, Czech Technical University in Prague, Brehova 7, Prague, Czech Republic*
[c] *CERN, 1211 Geneve 23, Switzerland*
[d] *University of Milano-Bicocca, Piazza dell'Ateneo Nuovo, 1, 20126, Milan, Italy*
[e] *University of Vienna, Universitätsring 1, 1010, Vienna, Austria*

A B S T R A C T

After the discovery of a cross-luminescence (CL) in BaF$_2$ in 1982, a large number of CL scintillators were investigated. However, no CL scintillator superior to BaF$_2$ has been discovered, and the research of CL scintillators has subsided. Recent technological development in medical imaging and high-energy physics created a new demand for ultra-fast scintillators further supported by the development of UV-sensitive semiconductor photodetectors. As a consequence, renewed interest in CL scintillators appeared. To satisfy the requirements of fast timing applications high photo-detection efficiency, e. i. a good spectral match between the scintillator and photodetector must be achieved. Cesium-based ternary chlorides could provide a red-shift (~1.5 eV) of CL towards the sensitive region of the photodetector (PMT or SiPM) while keeping light output and timing characteristics comparable to BaF$_2$.

## 1. Introduction

Nowadays ionizing radiation is used in almost all fields of human activity from geology, military, nuclear industry, environment protection, and medicine [1,2] and more and more demand for fast timing properties are requested. Medical imaging is one of the rapidly growing fields. Methods like single-photon emission tomography (SPECT) and positron emission tomography (PET) greatly improved cancer diagnostics and allowed more effective treatment of cancer which is the leading cause of death in modern society. Construction of the time-of-flight positron emission tomography (TOF-PET) scanner is the next important step in the field of medical imaging [3]. The introduction of TOF-PET is mainly motivated by two factors: an increase in signal-to-noise ratio (SNR) and the possibility to determine the position of the annihilation point along the line of response (LOR). Improving the coincidence time resolution (CTR) of the detection system, the ratio of SNR of the PET scanner with TOF and PET scanner without TOF is increasing according to:

$$\frac{SNR_{TOF\;PET}}{SNR_{PET}} = \sqrt{\frac{2 \cdot D}{c \cdot CTR}},\qquad(1)$$

where $D$ is the diameter of the field of view and $c$ is the speed of light in a vacuum [4,5]. Equation (1) shows that lowering CTR to 100 ps (FWHM) will result in an approximately 8-fold increase in the sensitivity and achieving 10 ps CTR will result in an approximately 25-fold increase in sensitivity of PET scanners. This would allow for a dramatic dose reduction in clinical practice. Moreover, achieving a CTR of 10 ps (FWHM) would allow determining position along LOR with an uncertainty of 1,5 mm [6]. Such spatial resolution is close to an inherent limitation of PET due to the positron mean free path in tissue. However, reaching the 10 ps benchmark is a challenging task and requires significant innovative technological advancements.

A scintillator together with a photodetector and readout electronics constitute the PET detector chain which sets a limit for the timing performance of the detector. Improvement of the scintillator timing characteristics is necessary to achieve CTR well below 100 ps. Several

---

concepts have been developed to either improve conventional scintillators or introduce new scintillators for fast timing applications (including PET). Various codoping strategies, direct gap semiconductor nanoscintillators, hot intraband luminescence, Cherenkov radiation [7], and cross-luminescence [8,9] have become of interest [10]. For example, a great improvement of timing characteristics of Pr doped silicates was achieved by codoping and detection of prompt photons [6, 11]. However, the best samples achieved CTR around 100 ps [12] which is still far from the 10 ps desired target. Taking into account the extensive development of the bulk silicate scintillators that led to the materials with such timing characteristics, a further breakthrough is implausible [3]. Hot intraband luminescence (IBL) features very fast decay times estimated to be of the order of 1 ps [13]. However, very low light yields up to 33 ph/MeV [14] and emission in the infrared region severely hinders the application potential of hot IBL. Detection of the prompt Cherenkov photons can be used to improve timing characteristics of conventional scintillators (e. g. YAG, LuAG, LYSO, or BGO) or for the construction of sole Cherenkov radiators. The concept of a sole Cherenkov radiator was first introduced by Ooba et al. [15]. Recently, a CTR of 71 ps was achieved using 5 mm thick $PbF_2$ crystal [16] and a CTR of 30 ps using 3.2 mm thick black lead glass [17] as a sole Cherenkov radiator. However, those results were obtained using a multichannel plate photomultiplier tube (MCP PMT) resulting in high costs and low detection efficiency. Direct-gap semiconductor nanostructures in the form of quantum dots, rods, and wells can feature very fast intense excitonic luminescence due to the quantum confinement effect. This was demonstrated on a wide variety of materials including ZnO:Ga [18], $CsPbBr_3$ [19], CdSe/CdS [20]. However, these powder materials suffer from poor light collection due to reabsorption and scattering. Moreover, the low stopping power of semiconductor nanoparticles limits their detection efficiency for 511 keV annihilation photons. This can be possibly solved via composite sampling pixels consisting of a fast nanoscintillator and a high $Z_{eff}$ scintillator [21]. However, this technology is at the very early stages of development and reabsorption remains a problem for such nanoscintillators due to their intrinsically small Stokes shift which is inherent to free Wannier exciton luminescence. Cross-luminescence (CL) is a very fast phenomenon that can be exploited for scintillation. It is observed mostly in cesium and barium-based halides. Significant research attention was devoted to CL, mainly in the 1990s [22]. However, most of the work was focused on basic research and no CL scintillator superior to barium fluoride ($BaF_2$) was discovered. $BaF_2$ is currently the only commercially available CL scintillator, but its CL emission within 180–220 nm is not compatible with commonly used alkali metal based photomultiplier tubes (PMTs). Numerical modeling [23,24] in cesium-based ternary chlorides confirms their UV positioned CL reported earlier in experimental works [25–28] better matching the sensitivity of alkali metal based PMTs. In theory, Cs-based cross-luminescence material could produce fast luminescence with sufficient light output for fast timing applications. Moreover, the single crystal form together with the large Stokes shift inherent to CL allows efficient light collection. However, CL materials suffer from low stopping power for high-energy photons due to low effective atomic number ($Z_{eff}$) and densities. Low stopping power results in low sensitivity, because of the requirement for small crystal dimensions in fast timing applications. Therefore, the development of CL scintillators involving heavy elements is necessary for the precise timing detection of high-energy photons. Some of the key material parameters including density, $Z_{eff}$, attenuation length, and melting point for $BaF_2$, $CsCaCl_3$, $Cs_2BaCl_4$, and $Bi_3Ge_4O_{12}$ (BGO) are listed in Table 1. The BGO was added as an example of a classical heavy scintillator.

In this work, we report a study of several novel Cs-based CL halide crystals. All reported crystals were grown by the vertical Bridgman method. Structural, optical, and scintillation properties of prepared crystals were studied using various techniques including X-ray diffraction, radioluminescence, photoluminescence, light yield, and scintillation decay measurements. The application potential is discussed.

## 2. Experimental

All reported crystals were prepared from starting materials of commercially available cesium chloride (CsCl 99.9%, Alfa Aesar), magnesium chloride ($MgCl_2$, 99.99% anhydrous, Sigma Aldrich), calcium chloride ($CaCl_2$, 99.99% ultradry, Alfa Aesar), strontium chloride ($SrCl_2$, 99.99% anhydrous, Sigma Aldrich) and barium chloride ($BaCl_2$, 99.99% anhydrous, Sigma Aldrich). All starting chemicals were treated by chlorinating agents according to Refs. [29,30] to remove oxidic anionic impurities. Furthermore, CsCl, $MgCl_2$, and $CaCl_2$ were purified by zone refining [31]. Our setup did not allow zone refining of $SrCl_2$ and $BaCl_2$ due to their high melting points. In the case of KCl and RbCl, their high purity as-grown single crystals were used as starting materials. All reported crystals were grown by a single-zone vertical Bridgman method using a micro-pulling-down apparatus. The starting charge (ca. 3–4 g depending on the specific material) was melted in quartz ampoule and subsequently pulled down with a rate of 0.6 mm/h in all performed growth experiments (see Fig. 1). For further details see Vanecek et al. [32]. All procedures consisting of quartz ampoules feeding and valve closure, handling and weighing of all chemicals, and manufacturing of grown crystals was performed in the atmosphere-controlled glovebox (MBraun Labstar) with the content of $O_2$ and $H_2O$ below 1 ppm.

For the X-ray powder diffraction (XRPD) analysis, the crystal samples were powdered in alumina mortar and pestle and placed in the Ø 0.5 mm borosilicate-glass capillary in the glovebox. The capillary was sealed with rubber to avoid air contamination. Powder diffraction data were collected using the Debye-Scherrer transmission configuration on the powder diffractometer Empyrean of PANalytical($\lambda Cu, K\alpha = 1.54184$ Å) that was equipped with a focusing mirror, capillary holder, and PIXcel3D detector.

Radioluminescence (RL) spectra measured in the spectral range of 190–800 nm at room temperature were obtained using a custom-made spectrofluorometer 5000 M, Horiba Jobin Yvon. Tungsten-cathode X-ray tube Seifert was used as the excitation source (at 40 kV, 15 mA). The detection part of the set-up consisted of a single grating monochromator and photon-counting detector TBX-04, Horiba IBH Scotland. Measured spectra were corrected for the spectral distortions. A routine spectrally unresolved scintillation decay was measured by means of fast photomultiplier (PMT) R7207-01, Hamamatsu working in the current regime, and Keysight InfiniiVision DSOX6002A digital oscilloscope where the sample was optically coupled directly to the PMT photocathode.

**Table 1**
Density, effective atomic number $Z_{eff}$, attenuation length for 511 keV photons $l_0$, and melting point $T_m$ of selected materials.

| Chemical formula | Density [g/cm³] | $Z_{eff}$ [-] | $l_0$ (at 511 keV) [mm] | $T_m$ [°C] |
|---|---|---|---|---|
| $BaF_2$ | 4.89 | 51 | 21 | 1368 |
| $CsCaCl_3$ | 2.95 | 42 | 37 | 910 |
| $Cs_2BaCl_4$ | 3.76 | 49 | 28 | 588 |
| $Bi_3Ge_4O_{12}$ | 7.13 | 71 | 10 | 1050 |

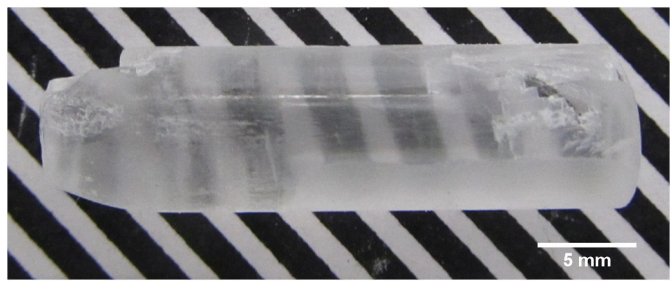

**Fig. 1.** As-grown crystal of $CsCaCl_3$ prepared by vertical Bridgman method.

Scintillation pulses were excited either by $^{137}$Cs γ-rays (662 keV) or $^{239}$Pu α-particles (5.15 MeV). The pulse-height spectra were collected by hybrid photomultiplier DEP PPO 475B, spectroscopy amplifier ORTEC 672 (shaping time set to 1 μs), and multichannel analyzer ORTEC 927TM. Scintillation pulses were excited by $^{137}$Cs γ-rays (662 keV). The crystals were covered with a thin film of UV/VIS-transparent Fluka immersion oil during all of the conducted measurements to prevent the degradation of the surface due to contact with atmospheric oxygen and moisture. This setup will be labeled "setup no. 1" in the text below.

In the case of the measurements of the fast scintillation decays with high resolution, the samples were excited by picosecond (ps) X-ray tube N5084 from Hamamatsu, operating at 40 kV. The X-ray tube is driven by the ps light pulser equipped with a laser diode operating at 405 nm. The repetition rate can go up to 10 MHz. The adjustable delay generator is triggering the laser pulses and the detector readout. The signal was detected by a hybrid picosecond photon detector HPPD-860 and Fluorohub unit (TCSPC method) from Horiba Scientific. The instrumental response function FWHM of the setup is about 76 ps. Samples are mounted a few centimeters in front of the beryllium window of the X-ray tube with a 45° angle to the incident beam. The luminescence is detected from the same surface by the detector. This setup will be labeled "setup no. 2" in the text below.

The scintillation time distribution of a CsCaCl$_3$ sample was measured using a TCSPC setup. A pulse diode laser from PicoQuant is used to excite an X-ray tube from Hamamatsu and produces an X-ray beam which is directed towards the crystal. The light generated by the sample excitation is then detected by a hybrid photomultiplier working in TCSPC mode. The sample under test was placed in such a way that the emitted light is detected from the same surface hit by the X-ray pulse. The instrumental response function of the whole setup is about 180 ps FWHM. The measurement was also repeated placing a 254 nm and a 300 nm optical bandpass filter with 40 nm FWHM in front of the hybrid PMT glass. This setup will be labeled "setup no. 3" in the text below.

The coincidence time resolution of a small 2 × 2 × 3 mm$^3$ CsCaCl$_3$ pixel was measured with the setup described in Ref. [33] against a known reference detector. The sample was wrapped in Teflon and Fluka Immersion oil was used between the VUV SiPM (3 × 3 mm$^2$ active area, Hamamatsu, S13370–3075CN) and crystal to remove the air gap and protect the crystal from air moisture. Only events depositing 511 keV both in the reference detector and the crystal under test were considered [33].

## 3. Results and discussion

### 3.1. Crystal growth and sample preparation

#### 3.1.1. (Cs,Rb)MeCl$_3$ (Me = Mg, Ca, Sr)

All as-grown crystals were extracted from the ampoule as a single ingot (See Fig. 1). The crystals were opaque due to the low quality of the surface, but their bulk was clear and transparent. Samples were cut and polished from as-grown crystals into transparent plates of thickness 1.5 mm for further optical evaluation (see Fig. 2).

X-ray powder diffraction measurements showed that CsMgCl$_3$ adopts the hexagonal crystal structure (space group $P6_3/mmc$, no. 194) isomorphous with CsNiCl$_3$ in accordance with literature [23,26,34]. While the CsCaCl$_3$ crystallizes in an undistorted simple perovskite structure (space group $Pm$-$3m$, no. 221). The most interesting behavior was observed in the case of CsSrCl$_3$. The diffractogram of CsSrCl$_3$ showed a major phase (84 %$_w$) with tetragonally distorted perovskite structure (space group $P4/mbm$, no. 127) and a minor phase (16 %$_w$) with orthogonally distorted perovskite structure (space group $Pnma$, no. 62). This was probably caused by successive phase transitions from cubic to tetragonal and from tetragonal to orthorhombic crystal structure occurring in CsSrCl$_3$ upon cooling [35,36] similarly to CsPbCl$_3$ [37]. These phase transitions should occur in the temperature region 90–120 °C [36]. Therefore we assume that the used cooling rate did not allow complete phase transformation from tetragonal to orthorhombic phase and the tetragonal phase "frozen" in the CsSrCl$_3$ sample. Taking into account that the sample was stored at room temperature for several weeks before the XRPD measurement. A similar phase transition was not observed in CsCaCl$_3$ as it occurs below room temperature ($T$ = 95 K) [38]. X-ray diffraction patterns of Cs$_x$Rb$_{1-x}$CaCl$_3$ (x = 0, 0.01, 0.05, 0.5, 1) showed (see Fig. 3) that for low Cs doping i. e. 0, 1, and 5 %$_{mol}$ the crystals adopted orthogonally distorted perovskite structure (space group $Pnma$, no. 62). However, a sample containing 50 %$_{mol}$ of Cs already adopts the same crystal structure as pure CsCaCl$_3$ (space group $Pm$-$3m$, no. 221) with reflections positions shifted towards higher angles. The shift is due to the smaller size of Rb$^+$ compared to Cs$^+$.

#### 3.1.2. A$_2$BaCl$_4$ (A = K, Rb, Cs)

All three materials from the A$_2$BaCl$_4$ matrix should, according to the literature [39–41], crystallize in cubic Th$_3$P$_4$ type crystal structure where A$^+$ and Ba$^{2+}$ occupy Th$^{4+}$ position in stoichiometric ratio 2:1 and Cl$^-$ occupies P$^{3-}$ position. However, detailed calorimetric studies revealed that these materials crystallize in a slightly off stoichiometric ratio namely K$_{2.08}$Ba$_{0.96}$Cl$_4$ [42] and Rb$_{2.07}$Ba$_{0.965}$Cl$_4$ [43]. Moreover, detailed phase equilibrium diagrams (PED) show the decomposition of the ternary compounds into constituents upon cooling [42]. This was omitted in Ref. [44] where the presence of CsCl in the grown crystals is confirmed by XRPD. Our observations support the decomposition of the ternary compound reported in Ref. [42] as crystalline ingots of all three A$_2$BaCl$_4$ compositions crumbled into powder upon cooling. However, several transparent irregularly shaped grains with dimensions of several mm remained in the ampoule even at room temperature. In the case of Rb$_2$BaCl$_4$, a large number of small grains were present in the ingot, which is the reason for the lower optical quality of the prepared sample. Cut and polished plates were prepared from the transparent grains for further optical characterizations (see Fig. 2).

X-ray powder diffraction of A$_2$BaCl$_4$ samples showed reflections of corresponding phases with cubic Th$_3$P$_4$ type crystal structure (space group $I4$-$3d$, no. 220). However, each reflection exhibits several satellites which could not be assigned to any record from the PDF4 2021 database. These reflections might be caused by a rhombohedral distortion of the cubic unit cell. Such an effect was proposed as an explanation for additional reflections in diffractograms of A$_2$BaCl$_4$ and A$_2$BaBr$_4$ powders [45].

### 3.2. Scintillation and luminescence characteristics

#### 3.2.1. (Cs,Rb)MeCl$_3$ (Me = Mg, Ca, Sr)

##### 3.2.1.1. Scintillation characteristics. Radioluminescence spectra of

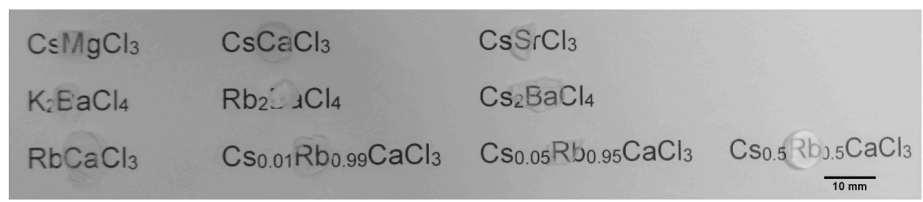

**Fig. 2.** Cut and polished samples.

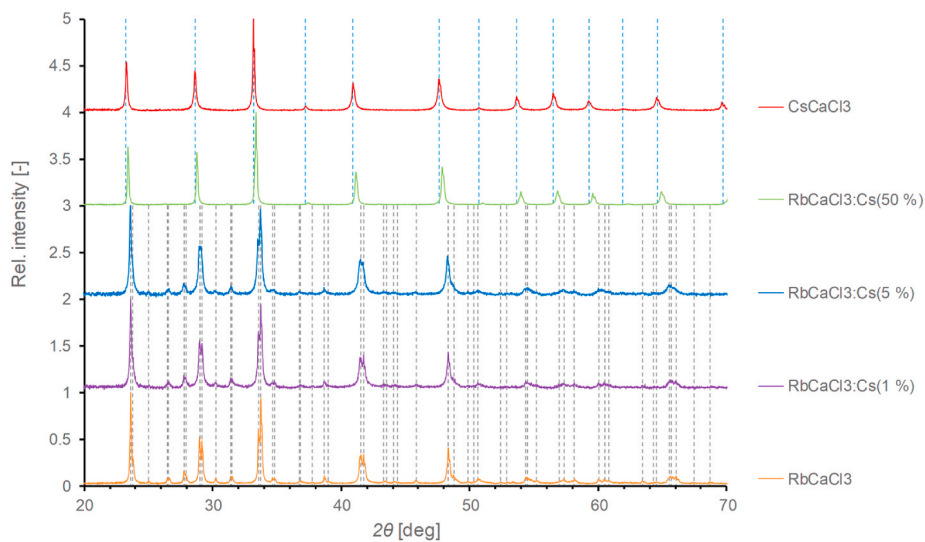

**Fig. 3.** Diffraction patterns of $Cs_xRb_{1-x}CaCl_3$ (x = 0, 0.01, 0.05, 0.5, 1) samples. Gray dashed lines show $RbCaCl_3$ (space group *Pnma*, no. 62) database record. Blue dashed lines show $CsCaCl_3$ (space group *Pm-3m*, no. 221) database record. (For interpretation of the references to colour in this figure legend, the reader is referred to the Web version of this article.)

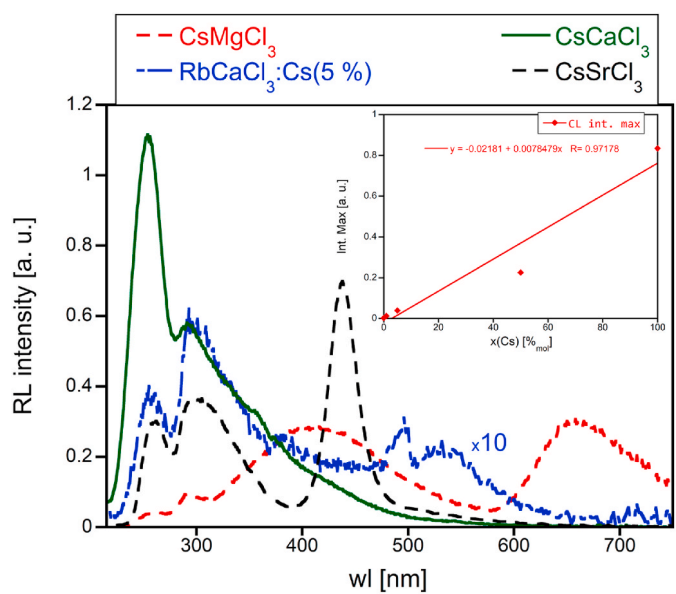

**Fig. 4.** RL spectra of (Cs,Rb)MeCl$_3$ (Me = Mg, Ca, Sr) crystals. The spectrum of RbCaCl$_3$:Cs(5 %$_{mol}$) was multiplied by 10 for better clarity. Inset shows the dependence of the intensity of high energy CL band with Cs concentration.

CsMeCl$_3$ samples are depicted in Fig. 4. The RL spectra show two main features: CL emission in the region 220–310 nm and exciton or defect-related emission at longer wavelengths. For all CsMeCl$_3$ samples, two emissions with maxima around 260 nm and 290 nm can be well observed in the high-energy region. This corroborates with the presumption that the CL in all reported materials originates from the radiative recombination between the hole in the outermost core band consistent of 5p states of Cs$^+$ and valence band mainly consistent of 3p states of Cl$^-$. Such presumption is supported by the results of band structure calculations [23,46]. Furthermore, positions of CL emission maxima are in good agreement with values reported in literature for CsCaCl$_3$ [25,46–48] and CsSrCl$_3$ [26,47,48]. The two-band structure of CL was theoretically predicted via cluster calculation for LiBaF$_3$ and KMgF$_3$ [47] which have the same coordination sphere (12 anion coordination with $O_h$ symmetry) of CL active cation. A more complex structure of the CL emission was reported for CsMgCl$_3$ [48] which should be the result of the lower symmetry ($C_3$) of the Cs coordination sphere. However, this was not observed in our study. Most probably due to both low intensity of CL in CsMgCl$_3$ and significant overlap with a broad emission band centered at 410. These effects obstruct the resolution of the complex structure of CL emission of CsMgCl$_3$. A similar overlap of the CL emission with excitonic or defect-related emission might be present in the CsSrCl$_3$ sample. Such overlap would explain the difference in width of bands centered at 260 nm and 290 nm is CsSrCl$_3$.

Luminescence in the lower-energy part of the RL spectra differs significantly among the CsMeCl$_3$ samples. The RL spectrum of CsMgCl$_3$ features two broad bands centered at 410 nm and 660 nm. The broad character suggests a defect or excitonic nature of those bands. In the RL spectrum of CsCaCl$_3$ no clear emission bands, besides the two ascribed to CL, can be distinguished. However, the long low energy tail of the RL spectrum suggests a contribution of luminescence of different origin that was tentatively ascribed to defect-related emission. In the case of CsSrCl$_3$, one relatively narrow band centered at 440 nm is observed. A very similar band was reported by Takahashi et al. in mixed crystals Cs$_{1-x}$Rb$_x$CaCl$_3$ [49] and CsCa$_{1-x}$Mg$_x$Cl$_3$ [50] and it was ascribed to self-trapped excitons (STE) luminescence in CsCaCl$_3$. However, this is in contradiction with our observation of this band in CsSrCl$_3$.

CL can be induced by Cs doping in CL-free alkali halides. The cesium impurity creates core levels below the valence band which meets the energy conditions for CL. The so-called impurity-induced CL was investigated in several binary [51–53] and ternary halides [46]. To investigate impurity-induced CL and to assess its potential for fast timing applications a series of Cs$_x$Rb$_{1-x}$CaCl$_3$ crystals were studied. Radio-luminescence spectra of two samples (x = 0.05 and 1) from the Cs$_x$Rb$_{1-x}$CaCl$_3$ series are presented in Fig. 4. The values in the spectrum for sample RbCaCl$_3$:Cs(5 %$_{mol}$) are multiplied by 10 for better clarity. All Cs containing Cs$_x$Rb$_{1-x}$CaCl$_3$ samples feature emissions centered at 260 nm and 290 nm which we ascribe to CL. The intensity of the CL emission increases monotonically with increasing Cs concentration. The difference in the intensity ratio of 260 nm and 290 nm emission among Cs$_x$Rb$_{1-x}$CaCl$_3$ samples is most probably due to the overlap of low energy CL emission (at 290 nm) with emission of different nature. Such emissions have most probably a different, if any, dependence of emission intensity on Cs content. Therefore, the intensity of the high energy emission (at 260 nm), which should be only very slightly perturbed, was used to examine the influence of Cs content on the intensity of CL in Cs$_x$Rb$_{1-x}$CaCl$_3$. The dependence of the intensity of the high energy CL emission (at 260 nm) with Cs concentration appears to be reasonably well fitted ($R^2 = 0.944$) with linear dependence (see inset of Fig. 4). The linear dependence of LY of fast emission on Cs concentration in $Cs_xRb_{1-x}CaCl_3$ was reported by Takahashi et al. [49]. However, the number of experimental points and their distribution within the examined concentration range in this study is not sufficient to make clear conclusions. The difference in RL intensity in the region approximately 320–540 nm is most probably due to defect-related luminescence. This assumption is based on the observation that RL intensity in this region is strongly dependent on the crystal quality. The RL spectra of the Cs free sample features only very weak (approx. 100 times less intense than BGO standard) overlying unresolved bands.

Scintillation decay profiles measured with setup no. 1 (see Fig. 5a) of all $CsMeCl_3$ samples can be well fitted by the convolution of the IRF with two exponential components. In the short time window (150 ns) a majority of the scintillation light (>99%) is emitted with decay times around 2 ns, which we ascribe to CL. However, increased background points toward the presence of slow components which were not resolved in the fast time window. Such slow components could be related to excitonic or defect emissions. A higher concentration of defects is to be expected for newly developed materials. Optimization of the crystal growth should result in the suppression of such defects. The contribution of slow components in $CsMeCl_3$ crystals was studied in a long time window (1500 ns) under excitation with α radiation of $^{239}$Pu. Since alpha radiation does not effectively excite CL [54,55] it is convenient for the study of the kinetics of slow scintillation components in CL scintillators. All three $CsMeCl_3$ show very similar scintillation kinetics under α irradiation. The majority (>97%) of the scintillation light is emitted with decay times around 40 ns. While the rest of the scintillation light is emitted with decay times from 200 to 400 ns. Parameters of the fit together with light yields of the $CsMeCl_3$ samples are summarized in Table 2.

Fast scintillation decay kinetics of $Cs_{1-x}Rb_xCaCl_3$ samples were studied using pico X-ray apparatus (setup no. 2). Fig. 5b depicts decay curves of samples containing 1, 5, and 100 %$_{mol}$ of Cs measured in the short time window (30 ns) together with mean decay times calculated from 2 exponential fitting of the decay curves. Acceleration of the scintillation decay with decreasing Cs content can be clearly seen. This effect could be explained by competition of CL and Auger recombination for the holes in the conduction band. However, this acceleration of scintillation decay is connected with a decrease in the intensity of CL emission.

From preliminary results, $CsCaCl_3$ emerged as the best candidate from materials in this study. Therefore, more detailed measurements were carried out to assess the potential of this material for fast timing applications. These involved measurements of fast scintillation decay (setup no. 3) and CTR. Fig. 6 depicts fast spectrally unresolved scintillation decays of $CsCaCl_3$. The decay curve is well fitted with the convolution of IRF and two exponential functions. First very fast component with a decay time of 151 ps and the second slower component with a decay time of 2.212 ns accounting for 6.47% and 93.53% of the scintillation light respectively. Moreover, one exponential component was used in the deconvolution to account for the scintillation rise time. The measured rise time was not resolvable within the resolution of the system and is most likely in the range of 0–50 ps.

Spectrally resolved scintillation decay curves were measured (setup no. 3) to further investigate scintillation kinetics in $CsCaCl_3$. Bandpass filters with center wavelengths (CWL) at 254 ± 20 and 300 ± 20 nm were used to separate the two emission bands (See Fig. 7). Scintillation decay curve measured with 250 nm CWL shows a single exponential decay of 2.158 ns. This value is in good agreement with the slower component in spectrally unresolved measurement. On the other hand, scintillation decay curve measured with 300 nm, CWL shows two exponential decay with values close to those for spectrally unresolved measurement ($t_{d1}$ = 266 ps, $t_{d2}$ = 2.144 ns). Therefore, we ascribe the ~2 ns component to CL in $CsCaCl_3$ and the ~200 ps component to a heavily quenched excitonic or defect-related emission.

Measurements of the CTR were challenging due to the hygroscopic nature of $CsCaCl_3$, for which the test bench was not designed. The transparency of employed oil was very low (below 280 nm) and only light above 280 nm was able to fully reach the SiPM, resulting in a significant light loss. Nevertheless, even with such unoptimized conditions, the measured CTR of $CsCaCl_3$ was 148 ± 12 ps FWHM, compared to 164 ± 12 ps obtained with $BaF_2$ measured under the same conditions of optical coupling and SIPM (though the emission of $BaF_2$ is deeper in the UV, suffering more from the coupling agent). Based on the measured scintillation properties of $CsCaCl_3$ and the first CTR measurements presented in this study, the expected timing performance should be similar to the one of $BaF_2$. With optimal conditions of wrapping, optical coupling, and SiPM (better PDE) a CTR close to 50 ps may be expected as obtained in $BaF_2$ [9].

*3.2.1.2. Photoluminescence characteristics.* The PL spectroscopy of $CsMgCl_3$ revealed the emission observed in the RL spectrum centered at 410 nm originates from at least two separate centers with their maxima at 356 and 460 nm and the corresponding excitation bands centered at 266 and 286 nm, respectively, see Fig. 8a. The decay kinetics of the center at 356 nm consists of a fast component with 1.2 and 5.6 ns day times and another component with 234 ns day time, while the decay of the center with emission at 460 nm has two components with 21 and 51 μs lifetimes, see Fig. 9. A complex structure of excitation bands can be seen below 250 nm. Excitation in this range leads to complex emission ranging from 250 to 650 nm. A similar emission structure can be seen under excitation below 250 nm either in the case of CsCaCl$_3$ and CsSrCl$_3$. In both cases, multiple emission bands span over the range 250–650 nm, and the spectra are dominated by the band with the maxima at 300 nm matching the emission observed in RL spectra. The spectra are not shown here for clarity. No other excitation bands were observed in CsCaCl$_3$ PL excitation spectra. The PL excitation spectrum of CaSrCl$_3$ with the emission set to the maximum of the narrow band at 434 nm seen in the RL spectrum reveals a pair of excitation bands with the maxima at 340 and 375 nm, see Fig. 8b. The decay kinetics measured with the emission fixed at 434 nm and excitation at 340 and 375 nm are dominated by 1.6 and 2.1 μs components, respectively. A slower component with a lifetime of about 6 μs is present in both cases.

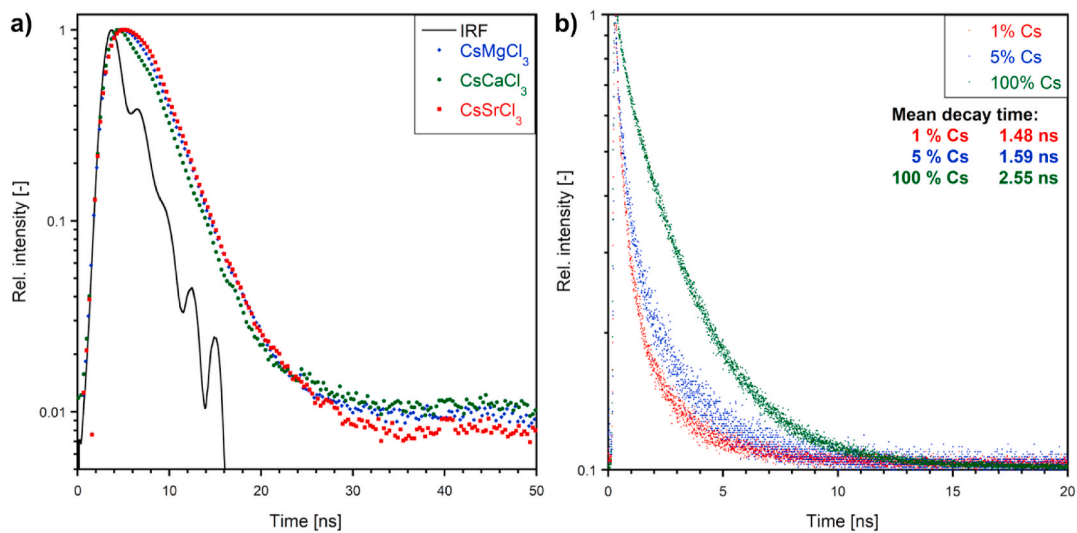

**Fig. 5.** Scintillation decay curves of a) $CsMeCl_3$ samples (setup no. 1) and b) fast scintillation decay curves of $Cs_{1-x}Rb_xCaCl_3$ samples (setup no. 2).

**Table 2**
Scintillation decay times $T_i$ and relative abundance $P_i$ of the $i$-th component of scintillation decay profile together with light yield for all presented materials.

| Excitation source: | $^{137}$Cs | | | | | $^{239}$Pu | | | | | |
| --- | --- | --- | --- | --- | --- | --- | --- | --- | --- | --- | --- |
| Composition | Ly [ph/MeV] | $T_1$ [ns] | $P_1$ [%] | $T_2$ [ns] | $P_2$ [%] | $T_1$ [ns] | $P_1$ [%] | $T_2$ [μs] | $P_2$ [%] | $T_3$ [μs] | $P_3$ [%] |
| CsMgCl$_3$ | 1113 | 2.36 | 99.7 | 178 | 0.3 | 43.4 | 97.3 | 0.402 | 2.7 | x | x |
| CsCaCl$_3$ | 1371 | 2.47 | 99.8 | 72.7 | 0.2 | 34.6 | 97.9 | 0.242 | 2.1 | x | x |
| CsSrCl$_3$ | 889 | 2.07 | 99.6 | 230 | 0.4 | 37.3 | 97.5 | 0.235 | 2.5 | x | x |
| K$_2$BaCl$_4$ | 817 | x | x | x | x | 76 | 95.2 | 2.04 | 4.4 | 30.0 | 0.4 |
| Rb$_2$BaCl$_4$ | 1181 | x | x | x | x | 52.9 | 90.6 | 2.83 | 9.0 | 29.9 | 0.4 |
| Cs$_2$BaCl$_4$ | 1369 | 1.68 | 99.3 | 140 | 0.7 | 50.9 | 96.7 | 2.20 | 2.8 | 27.4 | 0.5 |
| BGO (ref.) | 7311 | | | | | | | | | | |

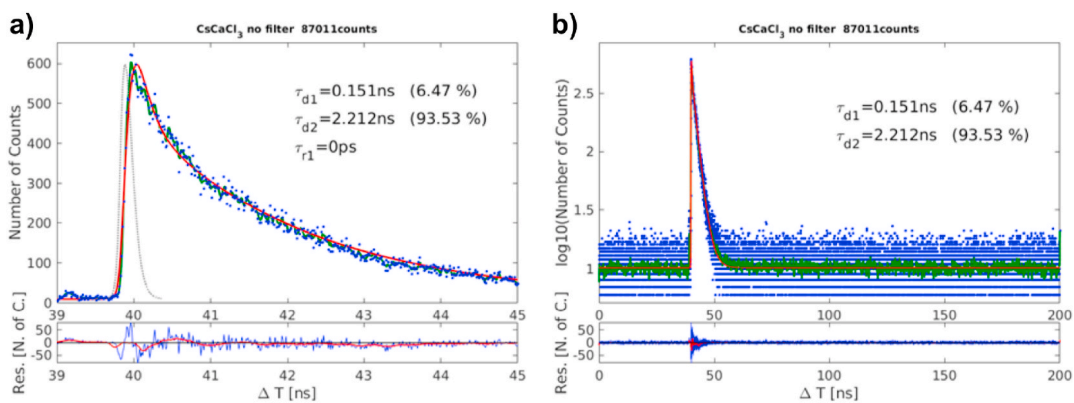

**Fig. 6.** Fast spectrally unresolved scintillation decays of CsCaCl$_3$ measured in a) short and b) long time window (setup no. 3).

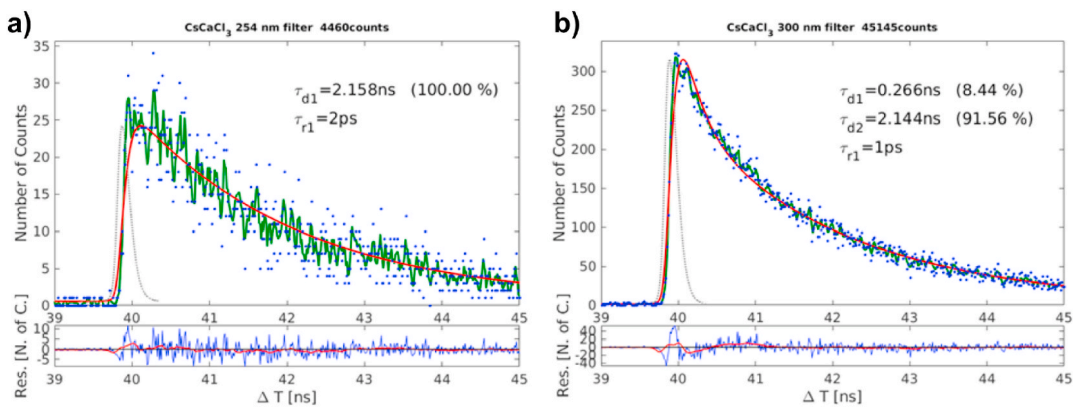

**Fig. 7.** Fast scintillation decays of CsCaCl$_3$ measured with a) 254 nm and b) 300 nm bandpass filters respectively (setup no. 3).

### 3.2.2. $A_2BaCl_4$ (A = K, Rb, Cs)

#### 3.2.2.1. Scintillation characteristics.
Fig. 10 depicts normalized RL spectra of the A$_2$BaCl$_4$ (A = K, Rb, Cs) samples. The inset of Fig. 10 shows the spectral dependence of the integral of RL intensity. All A$_2$BaCl$_4$ samples feature a very similar broad band centered at approx. 400 nm. The low energy side of this band shows a slight red-shift with an increasing atomic number of the A$^+$ kation. This suggests a similar nature of the emission center in all A$_2$BaCl$_4$ materials. Based on the position and width of this band we tentatively ascribe it to trapped excitons (TE). Furthermore, the Cs$_2$BaCl$_4$ sample features two bands centered at 260 nm and 290 nm respectively. The position of these bands is in good agreement with CL bands in CsMeCl$_3$ samples. This again corroborates with the presumption that CL in Cs$_2$BaCl$_4$ and CsMeCl$_3$ originates from the radiative recombination between the hole in the 5p state of Cs$^+$ and the electron in the 3p states of Cl$^-$. The difference in the intensity ratio of 260 nm and 290 nm emissions in CsCaCl$_3$ and Cs$_2$BaCl$_4$ is caused by the overlap of CL and TE emissions in Cs$_2$BaCl$_4$. The fact that we do not observe CL emission in $K_2BaCl_4$ and $Rb_2BaCl_4$ is in agreement with the results reported in the literature. On one hand, $K_2BaCl_4$ is not expected to exhibit CL since KCl does not satisfy the CL condition [47] i. e. the band gap is smaller than the energy difference between the uppermost core band and bottom of the valence band. On the other hand, the $Rb_2BaCl_4$ might satisfy the CL condition since it is satisfied in RbCl at low temperatures [47]. However, the CL in RbCl was observed only with very low intensity [56] due to significant self-absorption. Moreover, the CL emission in RbCl is reported at 190 nm which is at the very limit of our detection system. Both these facts prohibit the observation of CL in $Rb_2BaCl_4$ in our conditions.

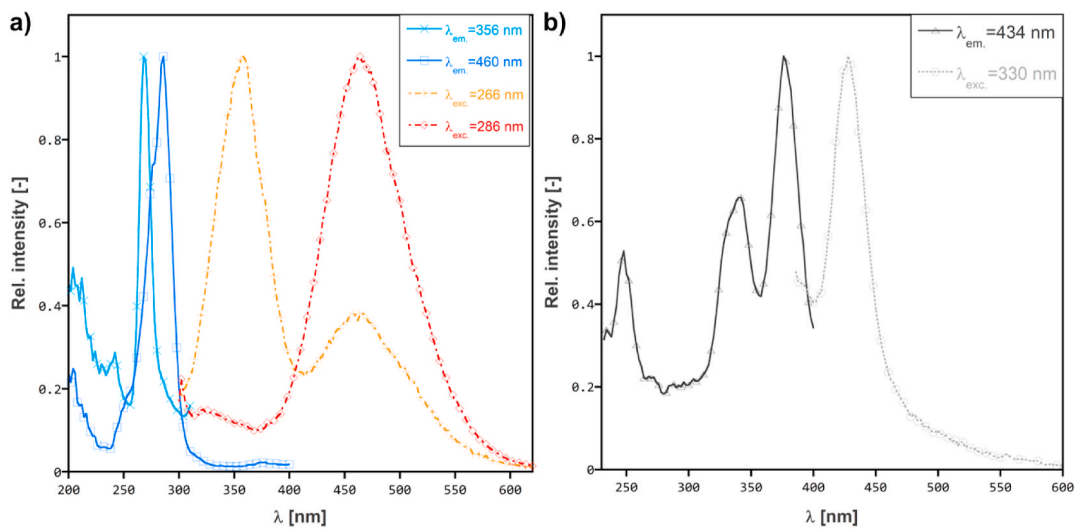

**Fig. 8.** Normalized PL and PL excitation spectra of a) $CsMgCl_3$ and b) $CsSrCl_3$ crystals.

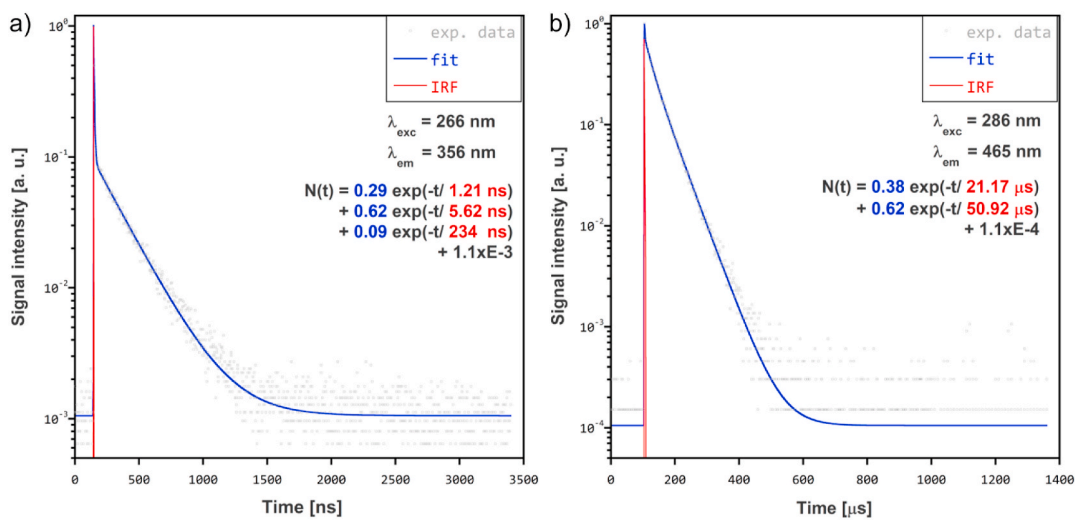

**Fig. 9.** PL decay curves of $CsMgCl_3$ excited with a) nanoLED and b) pulse Xe lamp.

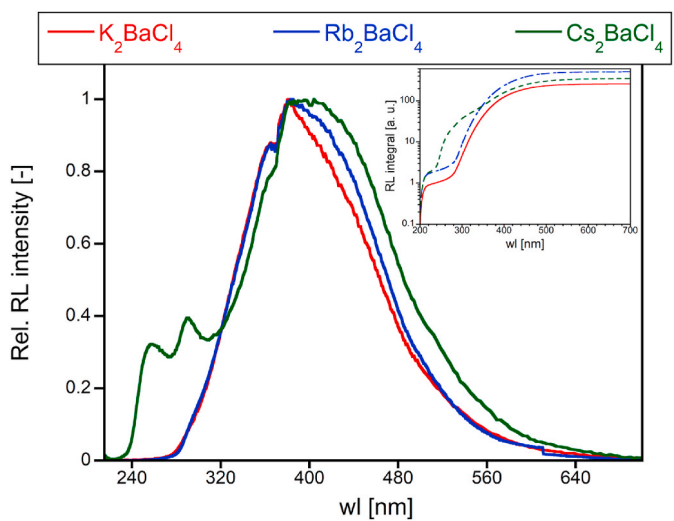

**Fig. 10.** Normalized RL spectra of $A_2BaCl_4$ crystals. The inset shows the spectral dependence of the integral of RL intensity.

Measurements of scintillation decay kinetics of the $A_2BaCl_4$ samples using setup no. 1 showed that only $Cs_2BaCl_4$ exhibits fast ∼ ns emission. This corroborates with observations from steady-state measurements above. Fig. 11a depicts the scintillation decay curve of the $Cs_2BaCl_4$ sample together with two exponential fit (convolution with IRF) and IRF. The timing characteristics are similar to $CsMeCl_3$ with the majority (>99%) of the scintillation light emitted with 1.68 ns day time. This supports the ascription of 250 nm and 290 nm bands in the RL spectrum of $Cs_2BaCl_4$ to CL. The slow components of the scintillation decay were examined with α excitation in the long time window (80 μs). Scintillation decay curves of all three $A_2BaCl_4$ samples (see Fig. 11b) were well fitted with three exponential components with similar decay times (see Tab 2). Short component with decay times 50–70 ns accounting for most of the scintillation light (>90%) and two slower components in the microsecond range with decays around 2.5 and 30 μs (See Table 2 for details). Increased contribution of the 2.83 μs component in $Rb_2BaCl_4$ could be caused by a higher concentration of the corresponding defect due to lower crystal quality compared to $K_2BaCl_4$ and $Cs_2BaCl_4$.

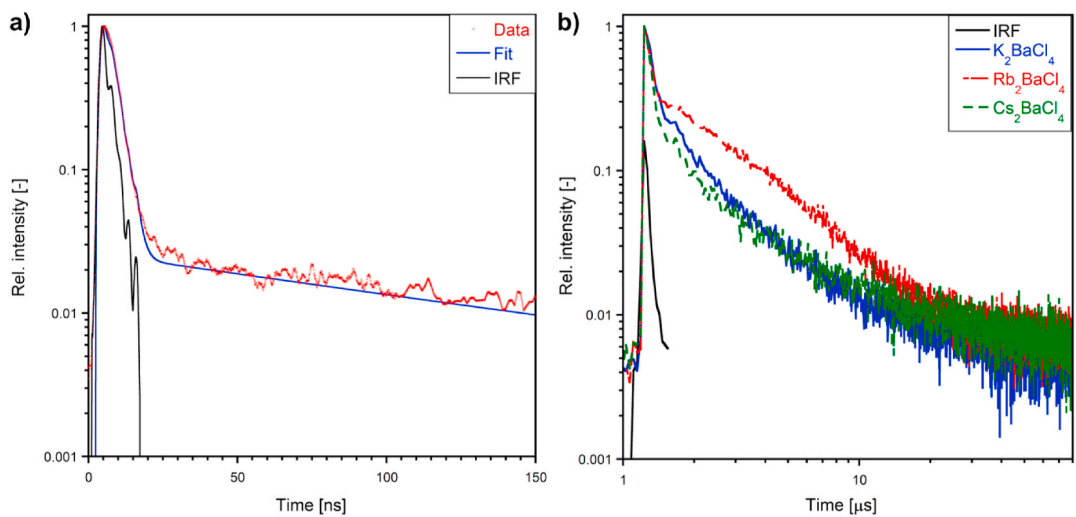

**Fig. 11.** Scintillation decay curves of a) $Cs_2BaCl_4$ excited with $^{137}Cs$ measured in the short time window and b) all three $A_2BaCl_4$ Samples excited with $^{239}Pu$ measured in the long time window.

*3.2.2.2. Photoluminescence characteristics.* The PL spectroscopy of $A_2BaCl_4$ samples reveals the complex structure of emission centers of these materials. Fig. 12a shows the PL excitation spectra of all examined $A_2BaCl_4$ samples measured with the emission wavelength fixed around the maxima of their RL spectra. A dominant excitation band centered at 245–250 nm is present in all samples, as well as a set of excitation bands above 260 nm. The complexity of the PL excitation spectra suggests the presence of multiple emission centers or an emission center with multiple configurations. Fig. 12b shows the PL spectra of the $A_2BaCl_4$ measured with excitation wavelength at about 250 nm. All the materials show a wide emission composed of multiple bands. The complexity of excited states is supported also by the measurement of PL decay kinetics. The PL decay kinetics of $Cs_2BaCl_4$ measured under excitation at about 250 nm and emission at 350 (360) nm in nanosecond and microsecond range are shown in Fig. 13. One can see the kinetics of the emission center is composed of several components with lifetimes ranging from 3 ns up to 18 μs. Analogous results were observed for $K_2BaCl_4$. In all of the examined materials, decay times were in the order of tens of microseconds for emission centers in the low-energy part of the spectra (above 400 nm).

## 4. Conclusions

Scintillation properties of perspective CL scintillators from $CsMeCl_3$ and $A_2BaCl_4$ families were investigated. All Cs containing samples exhibited two UV bands peaking at 260 nm and 290 nm with ns decay times which were ascribed to CL originating from the radiative recombination between the hole in the 5p state of $Cs^+$ and the electron in the 3p states of $Cl^-$. However, emission bands at longer wavelengths were present in all examined materials. Our results suggest that the emission bands at longer wavelengths originate from defects or excitonic effects. A high concentration of defects is expected for newly developed materials. However, optimization of technological processes (e. g. purification of starting materials, crystal growth, sample preparation, etc.) should result in suppression of such unwanted effects.

Impurity-induced CL was observed in $RbCaCl_3$ upon Cs doping ($Cs_{1-}$

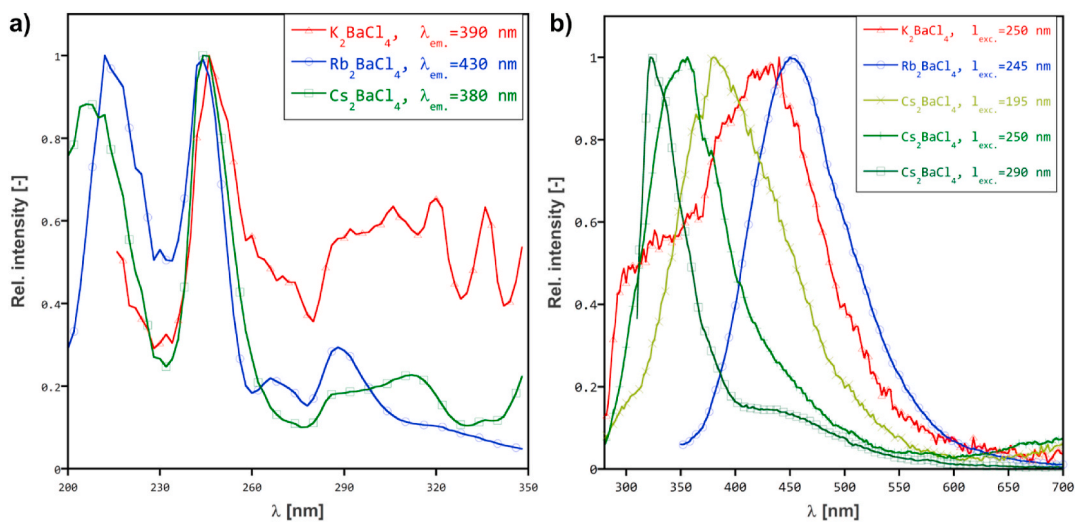

**Fig. 12.** Normalized a) PL excitation and b) PL spectra of $A_2BaCl_4$ crystals.

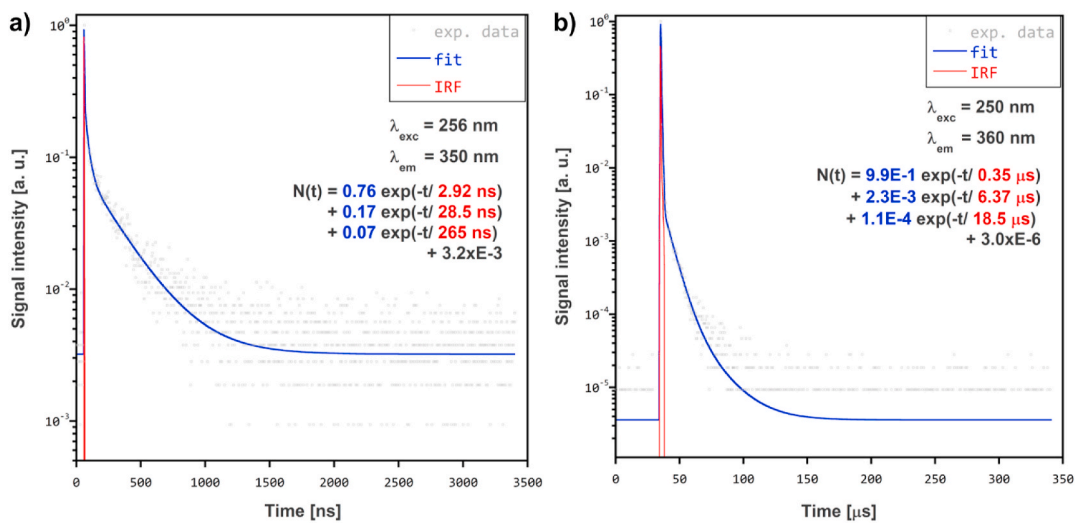

**Fig. 13.** PL decay curves of $Cs_2BaCl_4$ excited with a) nanoLED and b) pulse Xe lamp.

$_xRb_xCaCl_3$). We observed acceleration of the scintillation decay with decreasing Cs content. However, this acceleration is connected with a decrease in the intensity of CL. The intensity of the high-energy CL band decreases linearly with decreasing Cs concentration. Based on the reported results we suggest that this behavior is due to competition of Auger recombination (dominant in pure $RbCaCl_3$) and CL (dominant in $CsCaCl_3$) for the holes in the core band. This effect might be a perspective for achieving better CTR if the acceleration of scintillation decay kinetics would result in higher initial photon density despite the decrease in intensity due to Auger recombination.

From the preliminary results, the $CsCaCl_3$ emerged as the most promising candidate. The best measured samples featured scintillation decay with a mean decay time of 2.55 ns without a presence of slow and components LY close to 1400 ph/MeV. These results demonstrate the possibility of developing CL with properties comparable to $BaF_2$ and a better spectral match with the photodetector.

Further research should focus on the suppression of defects via optimization of the technological process of the prospective candidates and exploration of Cs based CL materials containing heavier elements. Such an effort could result in a Cs based CL scintillator that meets requirements for both high CTR and stopping power.

**Author contribution statement**

V. Vanecek: Conceptualization, Methodology, Writing - Original Draft, Investigation, Formal analysis, Visualization.

J. Paterek: Investigation, Formal analysis, Visualization, Writing - Review & Editing.

R. Kral: Conceptualization, Methodology, Investigation, Formal analysis, Writing - Review & Editing, Supervision.

R. Kučerkova: Investigation, Formal analysis, Visualization.

V. Babin: Investigation, Formal analysis, Visualization.

J. Rohlicek: Investigation, Formal analysis, Visualization.

R. Cala': Investigation, Formal analysis, Visualization, Writing - Review & Editing.

N. Kratochwil: Investigation, Formal analysis, Visualization, Writing - Review & Editing.

E. Auffray: Conceptualization, Supervision, Writing - Review & Editing, Project administration.

M.Nikl: Conceptualization, Writing - Review & Editing, Supervision, Project administration, Funding acquisition.

**Declaration of competing interest**

The authors declare that they have no known competing financial interests or personal relationships that could have appeared to influence the work reported in this paper.

**Acknowledgment**

This project was carried out in the framework of the Crystal Clear Collaboration. Support of project No. SOLID21 CZ.02.1.01/0.0/0.0/16_019/0000760) of the Operational Programme Research, Development and Education financed by European Structural and Investment Funds and the Czech Ministry of Education, Youth and Sports is gratefully acknowledged. The authors thank A. Bystricky and A. Cihlar for starting materials purification and preparation of growth ampoules.